\documentclass[9pt,final, twocolumn]{IEEEtran}

\let\labelindent\relax  %needed to use enumitem package

\usepackage{cite}
\usepackage{amsmath,amssymb,amsfonts}
\usepackage{graphicx,enumitem}
\usepackage{textcomp}
\usepackage{caption}
\usepackage[ruled,linesnumbered]{algorithm2e}  %,algochapter
\usepackage{algpseudocode}
\usepackage{acronym}
\usepackage[bookmarks,colorlinks]{hyperref}

\SetKwInput{KwData}{\textbf{Init}} 
\let\oldnl\nl% Store \nl in \oldnl
\newcommand{\nonl}{\renewcommand{\nl}{\let\nl\oldnl}}% Remove line number for one line

\acrodef{eeg}[EEG]{electroencephalogram}
\acrodef{ecog}[ECoG]{Electrocorticography}
\acrodef{hmm}[HMM]{Hidden Markov model}
\acrodef{dl}[DL]{deep learning}
\acrodef{dnn}[DNN]{deep neural network}
\acrodef{cnn}[CNN]{convolutional neural network}
\acrodef{dwt}[DWT]{discrete wavelet transform}
\acrodef{rnn}[RNN]{recurrent neural network}
\acrodef{flops}[FLOPs]{floating point operations}
\acrodef{lstm}[LSTM]{Long-Short Term Memory}
\acrodef{mi}[MI]{mutual information}
\acrodef{svm}[SVM]{support vector machine}
\acrodef{mical}[MICAL]{Mutual Information-based CNN-Aided Learned factor graphs}

\begin{document}
\title{MICAL: Mutual Information-Based CNN-Aided Learned Factor Graphs for Seizure Detection from EEG Signals}
% \author{First A. Author, \IEEEmembership{Fellow, IEEE}, Second B. Author, and Third C. Author, Jr., \IEEEmembership{Member, IEEE}
\author{B. Salafian$^{1}$, E. F. Ben-Knaan$^{2}$, N. Shlezinger$^{2}$,  S. Ribaupierre$^{3}$, and N. Farsad$^{1}$}

\author{Bahareh Salafian$^{1}$, Eyal Fishel Ben-Knaan$^{2}$, Nir Shlezinger$^{2}$, \\ Sandrine de Ribaupierre$^{3}$, and Nariman Farsad$^{1}$ % <-this % stops a space

\thanks{\textcolor{black}{Parts of this work were presented in the 2022 IEEE International Conference on Acoustics, Speech, and Signal Processing (ICASSP) as the paper~\cite{salafian2022cnn}.}
}
\thanks{
$^{1}$B. Salafian and N. Farsad are with the  Department of Computer Science at Ryerson University, Toronto, ON M5B 2K3
         {\tt\small \{bsalafian@ryerson.ca; nfarsad@ryerson.ca\}}}%
\thanks{$^{2}$E. F. Ben-Knaan and N. Shlezinger are with the School of Electrical and Computer Engineering, Ben-Gurion University of the Negev, Be'er Sheva, Israel, 84105
        {\tt\small \{eyalfish@post.bgu.ac.il; nirshl@bgu.ac.il\}}}% 
\thanks{$^{3}$S. de Ribaupierre is with Department of Clinical Neurological Sciences and the School of Biomedical Engineering, University of Western Ontario, ON N6A 5B9
        {\tt\small sderibau@uwo.ca}}%
     
        }
% \thanks{This paragraph of the first footnote will contain the date on 
% which you submitted your paper for review. It will also contain support 
% information, including sponsor and financial support acknowledgment. For 
% example, ``This work was supported in part by the U.S. Department of 
% Commerce under Grant BS123456.'' }
% \thanks{The next few paragraphs should contain 
% the authors' current affiliations, including current address and e-mail. For 
% example, F. A. Author is with the National Institute of Standards and 
% Technology, Boulder, CO 80305 USA (e-mail: author@boulder.nist.gov). }
% \thanks{S. B. Author, Jr., was with Rice University, Houston, TX 77005 USA. He is 
% now with the Department of Physics, Colorado State University, Fort Collins, 
% CO 80523 USA (e-mail: author@lamar.colostate.edu).}
% \thanks{T. C. Author is with 
% the Electrical Engineering Department, University of Colorado, Boulder, CO 
% 80309 USA, on leave from the National Research Institute for Metals, 
% Tsukuba, Japan (e-mail: author@nrim.go.jp).}}

\maketitle

\begin{abstract}
We develop a hybrid model-based data-driven seizure detection algorithm called Mutual Information-based CNN-Aided Learned factor graphs (MICAL) for detection of eclectic seizures from EEG signals. Our proposed method contains three main components: a neural mutual information (MI) estimator, 1D convolutional neural network (CNN), and factor graph inference. Since during seizure the electrical activity in one or more regions in the brain becomes correlated, we use neural MI estimators to measure inter-channel statistical dependence. We also design a 1D CNN to extract additional features from raw EEG signals. Since the soft estimates obtained as the combined features from the neural MI estimator and the CNN do not capture the temporal correlation between different EEG blocks, we use them not as estimates of the seizure state, but to compute the function nodes of a factor graph. The resulting factor graphs allows structured inference which  exploits the temporal correlation for further improving the detection performance. On public CHB-MIT database, We conduct three evaluation approaches using the  public CHB-MIT database, including 6-fold leave-four-patients-out cross-validation, all patient training; and per patient training. Our evaluations systematically demonstrate the impact of each element in MICAL through a complete ablation study and measuring six performance metrics. It is shown that the proposed method obtains state-of-the-art performance specifically in 6-fold leave-four-patients-out cross-validation and all patient training, demonstrating a superior generalizability.
\end{abstract}

\begin{IEEEkeywords}
Epilepsy, mutual information, factor graphs, convolutional neural network, deep learning, seizure, EEG, neural mutual information estimator.
\end{IEEEkeywords}

\section{Introduction}
\label{sec:introduction}
Epilepsy is a chronic neurological disorder that is accompanied by the sudden and unforeseen occurrence of signs or symptoms resulting from abnormal electrical activity in the brain that may cause seizures~\cite{fisher2005epileptic}. According to World Health Organization, about 50 million people worldwide are diagnosed with epilepsy~\cite{noauthor_epilepsy_nodate}. The extensive sudden discharges in neural brain activity due to epileptic seizures can lead to life-threatening impacts such as involuntary movements, sensations, and emotions and may cause a temporary loss of awareness and even death~\cite{sajobi2021quality}. 

\subsection{Diagnostic Tests for Epilepsy}
\label{sec:tools}
There are many different physiological tests as well as imaging and monitoring techniques used to evaluate if a person has a form of epilepsy, and the type of seizure the patient is experiencing. The physiological tests include reviewing medical history~\cite{malmgren2012differential}, performing blood tests to observe the metabolic or genetic disorders associated with the seizures~\cite{pal2010genetic,kandar2012epilepsy}, or monitoring other health conditions that could trigger epileptic seizures~\cite{kandar2012epilepsy}. Imaging and monitoring are the most popular tools for detecting epileptic seizures. In this regard, various screening techniques such as magnetoencephalogram (MEG)~\cite{van2019simultaneous}, computed tomography (CT)~\cite{salmenpera2005imaging}, positron emission tomography (PET)~\cite{spencer1994relative}, and magnetic resonance imaging (MRI)~\cite{kulaseharan2019identifying} have been employed. Among these techniques, \ac{eeg} is considered to be the most powerful method as it shows clear rhythmic electrical activities of the neurons~\cite{subasi_epileptic_2019}. 

There are two types of \ac{eeg} recordings: the invasive \ac{ecog}~\cite{yang2020seizure}, and the non-invasive scalp \ac{eeg}. \ac{ecog} is typically used when a patient is diagnosed with refractory surgery as it provides direct measurement of brain electrical activity by implanting electrodes on the cortex~\cite{van2014functional}. In scalp \ac{eeg}, multiple electrodes are placed on the scalp of individuals for recording electrical activity~\cite{yang2020seizure}. This technique is widely preferred as it is non-invasive, economical, and portable. %However, it suffers from high noise where the signal is not as clean as ECOG measurements. %In this work, we focus on scalp ac{eeg} and propose a novel algorithm for automatically detecting seizure events. 

From a clinical point of view, a neurologist can analyze abnormalities in \ac{eeg} signals through visual inspection to understand the presence or the type of epileptic seizures. However, this diagnosis is time-consuming as it requires careful inspection of data from long recording sessions by a neurologist~\cite{turk2019epilepsy}, and is subject to inter-observer variability~\cite{asif2020seizurenet}. Moreover, \ac{eeg} measurements are usually contaminated by undesired artifacts and noise that can interfere with neural information and cause misdiagnosis of epileptic seizures~\cite{jiang2019removal}. To address these issues, an automatic seizure detection algorithm is desirable.

\subsection{Automated Seizure Detection from EEG Signals} 
\label{sec:previous study}
%Many studies have been previously developed in automatic seizure detection and they are basically divided into feature-based and non-feature-based designs. We explain some of the prior works that are done in these areas in Subsections~\ref{sec:feature based}-\ref{sec:nonfeature based}. 

Several different methods for automatically detecting seizures from \ac{eeg} recordings have been proposed in the literature. These techniques can be categorized into two different approaches: Machine learning and signal processing approaches based on engineered features extracted from the recordings, and machine learning approaches applied to raw \ac{eeg} recordings. 

\subsubsection{Feature-Based  Detection}
\label{sec:feature based}
Spike detection is the most popular feature-based method that aims to identify seizure spikes in the multichannel \ac{eeg} recording with high sensitivity and selectivity~\cite{tzallas2012automated}. Generally, spike detection methods are divided into different categories including template matching~\cite{sankar1992automatic,vijayalakshmi2010spike}, mimetic analysis~\cite{gotman1976automatic,gotman1982automatic}, power spectral analysis~\cite{exarchos2006eeg}, wavelet analysis~\cite{indiradevi2008multi}, and techniques based on artificial neural networks~\cite{ko2000automatic,gabor1992automated,acir2004automatic}. 

Most of the recent feature-based designs use the extracted features as input to \ac{dl} algorithms for seizure detection. The authors in \cite{san2019classification} first generated three different features based on Fourier, wavelet, and empirical mode decomposition transforms. They then applied a shallow 2D \ac{cnn} and concluded that Fourier transform achieved the best results. Another wavelet-based deep learning approach was performed in \cite{akut2019wavelet}. In this method, \ac{dwt} was used to extract time-frequency domain features in five sub-band frequencies, and then, a 2D \ac{cnn} architecture was employed to learn the features from predefined coefficients. Applying 1D \ac{cnn} to time-frequency features was observed in \cite{chen2018cost,fukumori2019fully}. In the proposed method, first DWT of signals was processed, and a 1D \ac{cnn} architecture then performed detection.

Feature-based design mainly depends upon the expert definition of \ac{eeg} characteristics such as slop, duration, height, and sharpness, which are not sufficient enough to represent an epileptic seizure spike, and it can result in high false detection rate~\cite{wilson2002spike}. Moreover, using a different transforms such as Fourier or wavelet transform as input to \ac{dl} models requires careful engineering and considerable domain expertise to design a feature extractor that transforms the raw data into a suitable representation~\cite{lecun2015deep}. However, this strategy is difficult since various types of patterns appear when a seizure occurs~\cite{kim2020epileptic}. Moreover, many interfering artifacts in the signal, for example due to blinking or muscle activity, can have structures similar to the seizure patterns. 

\subsubsection{Signal-Based Detection}
\label{sec:nonfeature based}
In the past decade, different \ac{dl} models have been investigated and tested in the area of seizure detection and analysis of time series EEG signal. For example, in~\cite {johansen2016epileptiform}, after using a notch filter, a 1D \ac{cnn} with few convolutional layers was employed to detect interictal epileptiform spikes due to seizures. Acharya et al.~\cite{acharya2018deep} applied a 3-layer \ac{cnn} architecture to the normalized \ac{eeg} signals. A 2D \ac{cnn} architecture was implemented in \cite{liu2019deep} and applied to a multi-class classification problem, where the \ac{eeg} was labeled according to different stages of seizure. Boonyakitanont et al.~\cite{boonyakitanont_comparison_2019} proposed a detection scheme using raw \ac{eeg} records divided into 4-second blocks followed by a deep 2D \ac{cnn} architecture to learn the features from \ac{eeg} signals. They showed state-of-the-art performance based on the detection accuracy when per-patient training was employed. 

To capture the temporal dependencies, some prior work have explored \acp{rnn} \cite{vidyaratne2016deep,yao2019automated}. %These models usually encounter vanishing gradient when dealing with long-term dependent signals. Therefore, to overcome this limitation, recent studies are mostly considered \ac{lstm} and gated recurrent unit (GRU). 
Hussein et al.~\cite{hussein2018robust} used a deep \ac{rnn}, particularly an \ac{lstm} to the segmented \ac{eeg} signals to learn the most robust features from recordings. Aristizabal et al.~\cite{ahmedt2018deep} developed another \ac{lstm}-based seizure detection technique for six pairs of \ac{eeg} signals. The performance of GRU was explored in \cite{talathi2017deep}. In this model, the GRU-hidden unites were used to classify \ac{eeg} into three different classes: healthy, inter-ictal and ictal (i.e., seizure) states. Roy et al. \cite{roy2019chrononet} proposed an architecture termed ChronoNet by stacking multiple 1D convolution layers followed by GRU layers.

While signal-based designs are widely used  in the literature, training models that are generalizable and perform well across different patients requires large networks and very large datasets. This is due to patient-to-patient variability and the fact that \ac{eeg} recordings are inherently very noisy. Hence, end-to-end training on raw \ac{eeg} data may not achieve the best performance in practice due to lack of access to large datasets, as well as limitations imposed by computational complexity, both during training or during inference.  
The challenges associated with previous works motivate the formulation of a reliable automatic seizure detection algorithm which generalizes to different patients, benefits from both temporal and inter-channel correlation, and is computationally efficient facilitating its application in real-time.

\subsection{Contributions}
In this paper, we propose \ac{mical}, which is a hybrid model-based/data-driven approach \cite{shlezinger2020model} for automatic seizure detection. This algorithm contains three main novel aspects: 
\begin{itemize}
    \item In contrast to prior works, we carefully design a 1D \ac{cnn} to process the \ac{eeg} signals with a higher receptive field and minimal preprocessing. This results in extracting features that capture longer term dependencies from raw \ac{eeg} signals.
    
    \item We use a neural \ac{mi} estimator to compute the {\em inter-channel} dependence between \ac{eeg} channels during the seizure. When a seizure occurs in one or more \ac{eeg} channels, the patterns of other channel recordings are affected, and the signals between the channels become correlated at the beginning and during ictal (i.e., seizure) state~\cite{brazier1972spread,quintero2016new}. Compared to the traditional methods for evaluating levels of dependence, including cross-correlation, \ac{mi} can capture higher-order statistical dependence between recordings. This is helpful in seizure detection since non-linear relationships often exist between \ac{eeg} channels during seizure. 
    
    \item We propose an inference method that uses the estimates obtained from the extracted features at each time interval to form learned factor graph \cite{shlezinger2020inference} which captures the {\em temporal} correlations in \ac{eeg} recordings. By applying message passing over the learned factor graph, seizures can be detected in a computationally efficient manner compared to deep learning approaches based on \acp{rnn}. 
\end{itemize}

The performance of \ac{mical} is comprehensively explored via six performance measures as well as three evaluation strategies. Using an extensive ablation study we systematically show that each component of \ac{mical} contributes positively to its performance. Comparing \ac{mical} with prior works, we demonstrate that our proposed method achieves state-of-the-art results in terms of performance and generalizability,  with the gains being most notable when given access to medium sized datasets.

\subsection{Organization}

The rest of this paper is organized as follows. In Section~\ref{sec:problem}, we discuss the seizure detection problem as well as the challenges related to traditional approaches for seizure detection. Section~\ref{sec:motivation} describes the proposed \ac{mical}. A comprehensive numerical evaluation of \ac{mical} is reported in Section~\ref{sec:eval}. Finally, we conclude the paper with summary discussions and future research direction in Section~\ref{sec:conclusion}.

\section{EEG-Based Seizure Detection System Model}
% \textcolor{red}{This section needs work. You need to define the problem statement using a similar approach as in ICASSP, i.e., using a mathematical formulation. The rest of the section is like a survey of prior work. Please note that this is not a survey paper, and this level of detail is not needed. You need to summarize Sub-Sections A, B and C in a few paragraphs. }

\label{sec:problem}
%As mentioned in Subsection~\ref{sec:tools}, \ac{eeg} is one of the most essential tools used in epileptic seizure studies as it shows the details about the electrical brain activity. 
The traditional way to analyze \ac{eeg} signals for seizure detection is to visually monitor the recordings by an expert. However, visual reviewing is time-consuming, and may results in a non-objective analysis with non-reproducible results. In addition, this process is usually accompanied by human errors due to, e.g., the subjective nature of the analysis, ictal (i.e., the seizure) spikes morphology, and the similarity of these signals to the waves or artifacts during normal operation of the brain. Therefore, an automatic seizure detection system is desirable, as it can enable long-term patient monitoring, reduce diagnosis time, and help neurologists to select the best treatment options for patients with epileptic seizures. An additional motivating factor stems from that fact that, in some cases, patients or their families are asked to report the number of seizures that occur during their daily lives. However, this approach has considerable limitations due to false descriptions of seizures and their frequency. Hence, automatic seizure detection provides a more elaborate and accurate technique for quantifying the number seizures, which would be helpful in research, diagnosis, and selecting the proper treatment option. %This has motivated the development of automatic seizure detection techniques. %Previous detection techniques are sometimes based on manual feature extraction by neurologists, which does not work well for a larger population. Besides, existing machine learning and \ac{dl} models are not able to capture both temporal and spatial correlations effectively and efficiently. We describe some of the prior studies in Section~\ref{sec:previous study}.      

In this paper, seizure detection refers to the identification and localization of the ictal time intervals from EEG recordings of patients with epileptic seizures~\cite{emmady_eeg_2020}. To formulate this mathematically, let $\boldsymbol{X}=\{\boldsymbol{X}_{1},\boldsymbol{X}_{2},\cdots,\boldsymbol{X}_{N} \}$ be the EEG recordings of a patient, where $N$ represents the number of channels. Each measured channel $\boldsymbol{X}_i$ is comprised of $n$ consecutive blocks, e.g., blocks of 1-second recordings, and we write $\boldsymbol{X}_i=[\boldsymbol{x}^{(i)}_{t_1},\boldsymbol{x}^{(i)}_{t_2},\cdots,\boldsymbol{x}^{(i)}_{t_n}]$, where $\boldsymbol{x}^{(i)}_{t}$ is the signal corresponding to the $i$-th \ac{eeg} channel during the $t$-th block. The seizure state for each block is represented as a binary vector $\boldsymbol{s}=[s_{t_1},\ldots s_{t_n}]$, where $s_t\in\{0,1\}$ models whether or not a seizure occurs in the $t$-th block. Our goal is to design a system that maps the \ac{eeg} recordings $\boldsymbol{X}$ into an estimate of $\boldsymbol{s}$, which is equivalent to finding the time indices where seizure occurs.  

\begin{figure}[!t]
  \centering
  \captionsetup{justification=centering}
\centerline{\includegraphics[width=0.9\columnwidth]{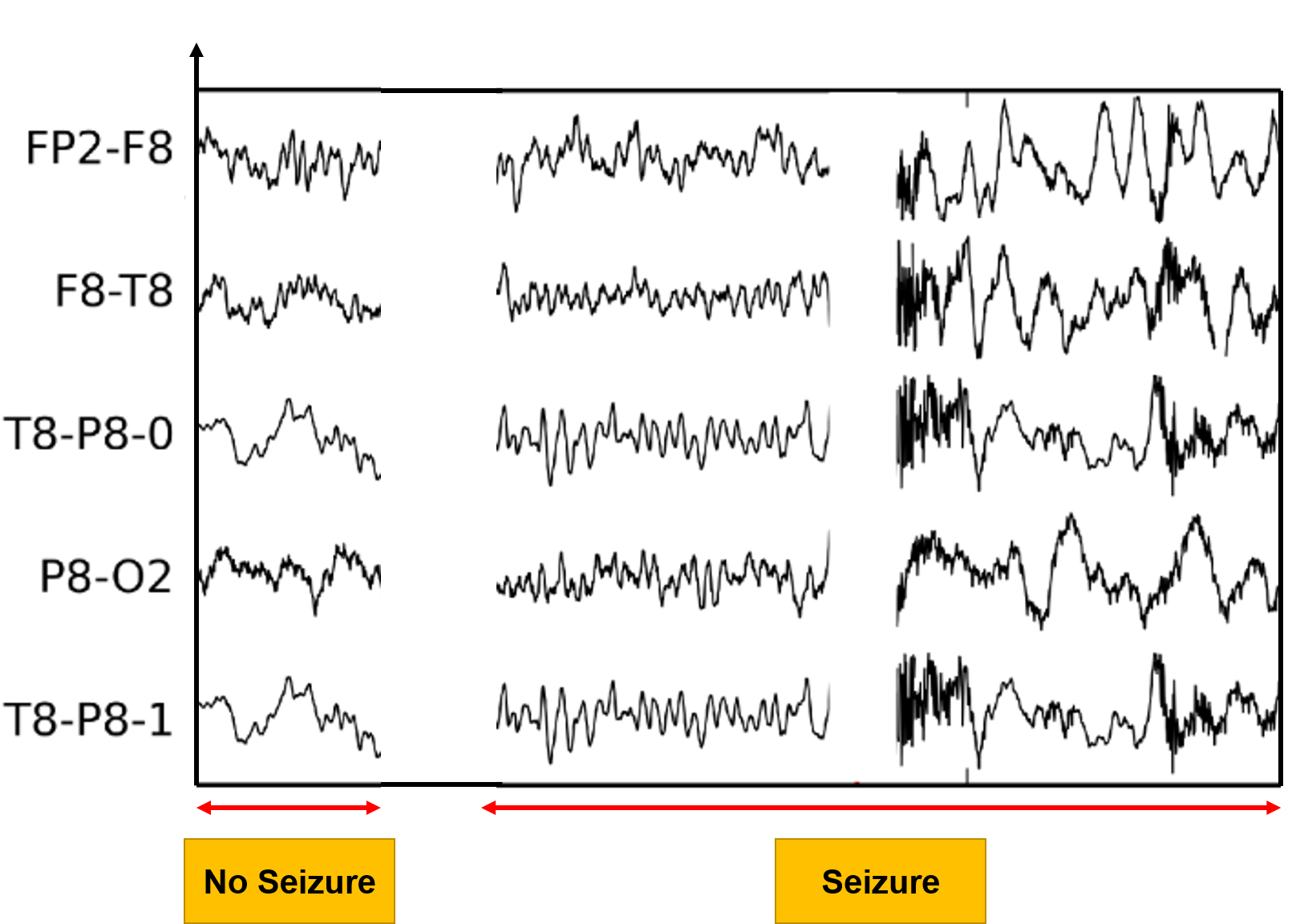}}
\caption{Inter-channel correlation during seizure vs. no-seizure~\cite{jemal2021study}.}
\label{fig:correlation}
\end{figure}

To model the relationship between the \ac{eeg} signals $\boldsymbol{X}$ and the seizure states $\boldsymbol{s}$, one must consider both inter-channel dependence as well as temporal correlations underlying the recordings. The former stems from the fact that when the seizure starts, the epileptic activity propagates to other areas in the brain~\cite{quintero2016new}, which affects the patterns of other channel recordings~\cite{jemal2021study}. % as illustrated in Fig.~\ref{fig:correlation}. 
This manifests as a high dependence between different channels, i.e., between $\boldsymbol{x}^{(i)}_{t}$ and $\boldsymbol{x}^{(j)}_{t}$, $i \neq j$, when $t$ is at the beginning and during ictal phase. Fig.~\ref{fig:correlation} demonstrates the signal patterns during seizure vs. no-seizure states. Our proposed solution, detailed in Subsection~\ref{sec:MI_estimation}, uses neural \ac{mi} estimators to capture this dependency.

Temporal correlation results from the fact that seizures typically span multiple recording blocks. Thus, the probability of observing a seizure at time instance $t$ depends on the presence of a seizure in the previous block, and as a result the entries of $\boldsymbol{s}$ can be approximated by a Markovian structure \cite{lee_classification_2018}. Our proposed solution, detailed in Subsection~\ref{sec:FG}, exploits this statistical structure using factor graphs. 

In the next section, we describe our proposed approach to automatic seizure detection in detail.

\section{Proposed \ac{mical} Algorithm}
\label{sec:motivation}
In this section we present the proposed \ac{mical} seizure detector. Our design of \ac{mical} is based on the following considerations:
\begin{enumerate}[label={\em C\arabic*}]
    \item \label{itm:MI} The level of statistical dependence between different channels provides an indication for the presence of a seizure.  
    \item \label{itm:RawSignal} Direct processing of the signal is preferable as it avoids the need for careful feature engineering. 
    \item \label{itm:Temporal} The temporal correlation between different blocks can be approximated as obeying a Markovian structure.
    \item \label{itm:Complexity} The detection algorithm must be operable with low complexity and should not require massive data sets for its training.
\end{enumerate}

% TODO explain how these are taken into account in MICAL
% These challenges associated with previous works motivate us to introduce a hybrid model-based/data-driven approach that is capable of capturing temporal correlation efficiently as well as the inter-channel correlation among \ac{eeg} channels during the seizure. As we will show using extensive numerical evaluations, this approach results in an algorithm that can achieve state-of-the-art detection performance and better patient-to-patient generalizabity compared to prior work, especially over medium sized datasets.

\begin{figure*}
    \centering
    \includegraphics[width=\linewidth]{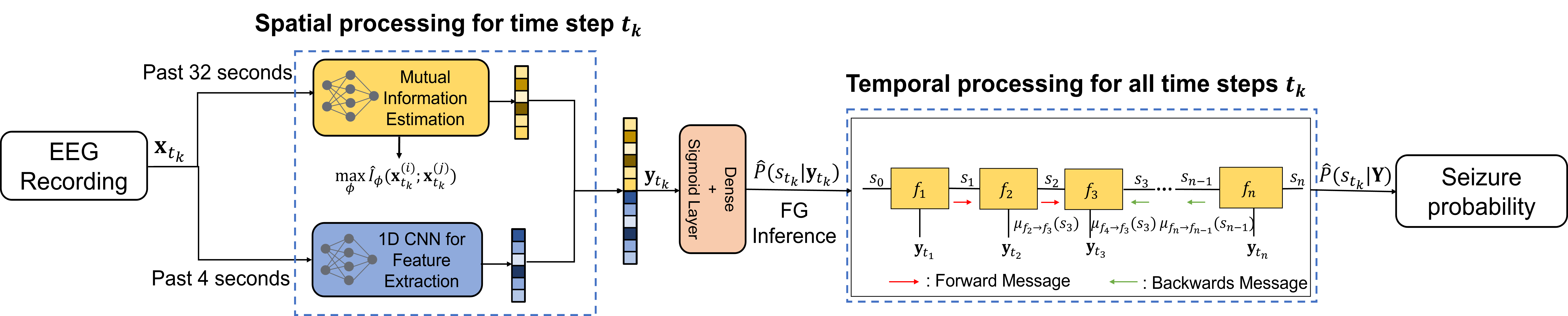}
    \caption{\ac{mical} illustration including spatial processing via \ac{mi} estimation and 1D \ac{cnn} processing followed by factor graph inference for temporal processing}
    \label{fig:method}
\end{figure*}

% The inter-channel correlation occurs due to the impact of seizures on electrical brain activity. When a seizure occurs in one or more \ac{eeg} channels, the patterns of other channel recordings are affected, and the signals between the channels become correlated at the beginning and during ictal (i.e., seizure) state~\cite{brazier1972spread,quintero2016new}. Fig.~\ref{fig:correlation} demonstrates the signal patterns during seizure vs. no-seizure states. Similarly, temporal correlations occur simply because seizure typically span across multiple blocks. 
Based on these consideration, we propose \ac{mical}, whose structure is illustrated in Fig~\ref{fig:method}. In the rest of this section, we describe each component of \ac{mical}. To account for the inter-channel dependence \ref{itm:MI},  we employ a neural \ac{mi} estimator block described in Subsection \ref{sec:MI_estimation}; To extract more  features from the \ac{eeg} recordings directly following \ref{itm:RawSignal}, we design a 1D \ac{cnn} architecture detailed in Subsection~\ref{sec:1DCNN}. These two blocks are used to estimate seizure probability for a given block. Finally, to account for temporal correlations following \ref{itm:Temporal} and do so in computationally efficient manner \ref{itm:Complexity}, we use estimates over different blocks to form a learned factor graph, as detailed in Subsection~\ref{sec:FG}.  

% \begin{figure*}

% \begin{minipage}[b]{1.1\linewidth}
%   \centering
%   \captionsetup{justification=centering}
%   \centerline{\includegraphics[width=17cm]{Figures/Diagram_test.png}}
% %  \vspace{2.0cm}
% %   \centerline{(a)Dependence among signals during seizure vs. no-seizure}\medskip
% \end{minipage}
% %
% \caption{\ac{mical} illustration.}
% %\textcolor{red}{change the neural network to the SMILE block. It is not clear from the diagram how is the output of SMILE combined with the output of the CNN. And shouldn't there be a softmax layer at the output?}

% \label{fig:method}
% %
% \end{figure*}

\subsection{Neural Mutual Information Estimation}
\label{sec:MI_estimation}

In order to compute inter-channel correlation among recordings, the most popular approach is cross-correlation, which is a measure of similarity between one signal and the time-delayed version of other signals. However, this method cannot capture the nonlinear relationship between samples, which are likely to occur in \ac{eeg} signals during the seizure. Unlike cross-correlation,~\ac{mi} represents higher-order joint statistics and is thus able to capture arbitrary statistical dependence between samples, even in the presence of nonlinear relationship between the signals. Mathematically, \ac{mi} can be formulated as:
% eq.1 in belghazi paper
\begin{equation}
  \label{eqn:mi}
I(X_1;X_2) = \int_{\mathcal{X}_1\times \mathcal{X}_2}\text{log}\left( \dfrac{d\mathbb{P}_{X_1X_2}}{d\mathbb{P}_{X_1} \otimes \mathbb{P}_{X_2}}\right) d\mathbb{P}_{X_1X_2}
\end{equation} 
where $\mathbb{P}_{X_1X_2}$ is the joint probability distribution and $\mathbb{P}_{X_1}$ and $\mathbb{P}_{X_2}$ are marginals. The $X_1$ and $X_2$ represent two random variables where in the case of seizure detection, they can be interpreted as the recordings for two different channels. 

The \ac{mi} can be expressed as the Kullback-Leibler (KL) divergence between the joint and the product of the marginals of two random variables $X_1$ and $X_2$~\cite{kullback1997information}:
\begin{equation}
  \label{eqn:kl diverge}
I(X_1,X_2) = D_{KL}(\mathbb{P}_{X_1X_2}\Vert \mathbb{P}_{X_1} \otimes \mathbb{P}_{X_2} )
\end{equation} 
where $D_{KL}$ is defined as:
\begin{equation}
  \label{eqn:kl diverge}
D_{KL}(\mathbb{P} \Vert \mathbb{Q}):=\mathbb{E}_\mathbb{P}\left[\log\frac{d\mathbb{P}}{d\mathbb{Q}}\right]
\end{equation}

Although \ac{mi} is a reliable measure to capture statistical dependence, the exact calculation based on \eqref{eqn:mi} and \eqref{eqn:kl diverge} for finite continuous and non-parametric \ac{eeg} samples is challenging~\cite{paninski2003estimation}. To facilitate \ac{mi} computation we use the Smoothed Mutual Information “Lower-bound” Estimator (SMILE) of \cite{song2019understanding}, which provides further improvements on Mutual Information Neural Estimator (MINE) proposed in~\cite{belghazi2018mutual}. Thus, to describe our \ac{mi} estimator, we briefly explain the operation of MINE and that of SMILE.  

A key technical aspect of MINE is dual representations of the KL-divergence, which is based on Donsker-Varadhan representation~\cite{donsker1975asymptotic}. This representation leads to the following lower bound where the supremum is taken over all functions $T$ such that the two expectations are finite.  
\begin{equation}
  \label{eqn:dual kl diverge}
D_{KL}(\mathbb{P} \Vert \mathbb{Q})\ge\sup_{T \in \mathcal{F}}\mathbb{E}_{\mathbb{P}}\left[T\right]-\log\left(\mathbb{E}_{\mathbb{Q}}\left[e^{T}\right]\right)
\end{equation}
Using both \eqref{eqn:kl diverge} and dual representation of KL-divergence, the idea is to choose $\mathcal{F}$ to be the set of functions ${T_\theta}: \mathcal{X}_1\times\mathcal{X}_2 \xrightarrow{}\mathbb{R}$ parametrized
by a deep neural network with parameters $\theta \in \Theta$, and the apply moving average gradient ascent to find the optimal parameters. This network is called statistics network, where the bound is calculated as: 
\begin{equation}
  \label{eqn:mi network}
I(X_1;X_2)\ge I_\theta(X_1,X_2)
\end{equation}
and the neural information measure, ${I_\theta(X_1,X_2)}$ is defined as:
\begin{equation}
  \label{eqn:NN information measure}
I_\theta(X_1,X_2)=\sup_{\theta \in \Theta}\mathbb{E}_{\mathbb{P}_{X_1X_2}}\left[T_\theta\right]-\log\left(\mathbb{E}_{\mathbb{P}_{X_1}\otimes\mathbb{P}_{X_2}}\left[e^{T_\theta}\right]\right)
\end{equation}

An important limitation of MINE is the large variance of the estimator, which can grow exponentially with the ground truth \ac{mi} value to be estimated from the samples. In order to solve the variance problem,~\cite{song2019understanding} proposes SMILE by introducing a clipping function in \eqref{eqn:NN information measure}, resulting in
% by considering $T$ as the invertible function of the density ratio. To achieve this purpose, the variational $f$-divergence proposed in.~\cite{nguyen2010estimating} is used, leading to the following lower-bound for~\ac{mi} estimation: 
% \textcolor{red}{I don't understand this description of how SMILE is obtained. There are several notations here which are not defined, and it is not clear why do we need to introduce this new notation of $I_{NWJ}$}
% (Nyugen et al. (NWJ)). ${\forall P,Q} \in \mathcal{P(X)}$ such that $P\ll\ Q$,
% \begin{align}
% &D_{KL}(\mathbb{P} \Vert \mathbb{Q}) \notag \\
% &=\sup_{T \in L{^\infty}(Q)}
% \left\{\mathbb{E}_{\mathbb{P}}\left[T\right]-\log\left(\mathbb{E}_{\mathbb{Q}}\left[e^{T-1}\right]\right):=I_{NWJ}(T)\right\},
%   \label{eqn:smile kl}
% \end{align}
% and $D_{KL}(\mathbb{P} \Vert \mathbb{Q})=I_{NWJ}(T)$ when $T=\log{\frac{dP}{dQ}}+1$ which leads to unbiased mini-batch estimator. 
% To address the issue of high variance, the partition function is clipped and the following estimator with smooth partition function is achieved~\cite{song2019understanding}:
% %
% % \begin{equation}
% %   \label{eqn:smile}
% % I_{SMILE}(T_\theta,\tau)=\mathbb{E}_{\mathbb{P}}\left[T_\theta(X_1,X_2)\right]-\log\mathbb{E}_{\mathbb{Q}}\left[clip(e^{T_\theta(X_1,X_2)},e^{-\tau},e^{\tau})\right]
% % \end{equation}
% %
\begin{align}
\hat{I}_{\boldsymbol{\theta}}(X_1; X_2)=& \mathbb{E}_{P_{{X_1}X_2}}\left[T_{\boldsymbol{\theta}}\right]\notag \\ 
&\qquad -\log\mathbb{E}_{P_{X_1} P_{X_2}}\left[{\rm clip}(e^{T_{\boldsymbol{\theta}}},e^{-\tau},e^{\tau})\right],
  \label{eqn:smile}
\end{align}
where ${\rm clip}(v,l,u)=\max(\min(v,u),l)$, and $\tau$ is a constant parameter that provides a knob for better tuning the bias-variance trade-off. In \cite{song2019understanding}, it was shown that while $\tau$ can be tuned to reduce the variance, it may not increase the bias significantly. %Therefore, in SMILE \cite{song2019understanding}, $T_\theta$ is a neural network that estimates the log-density ratio with hyper-parameter $\tau$. 

\begin{figure*} 
  \centering
  \centerline{\includegraphics[width=17cm]{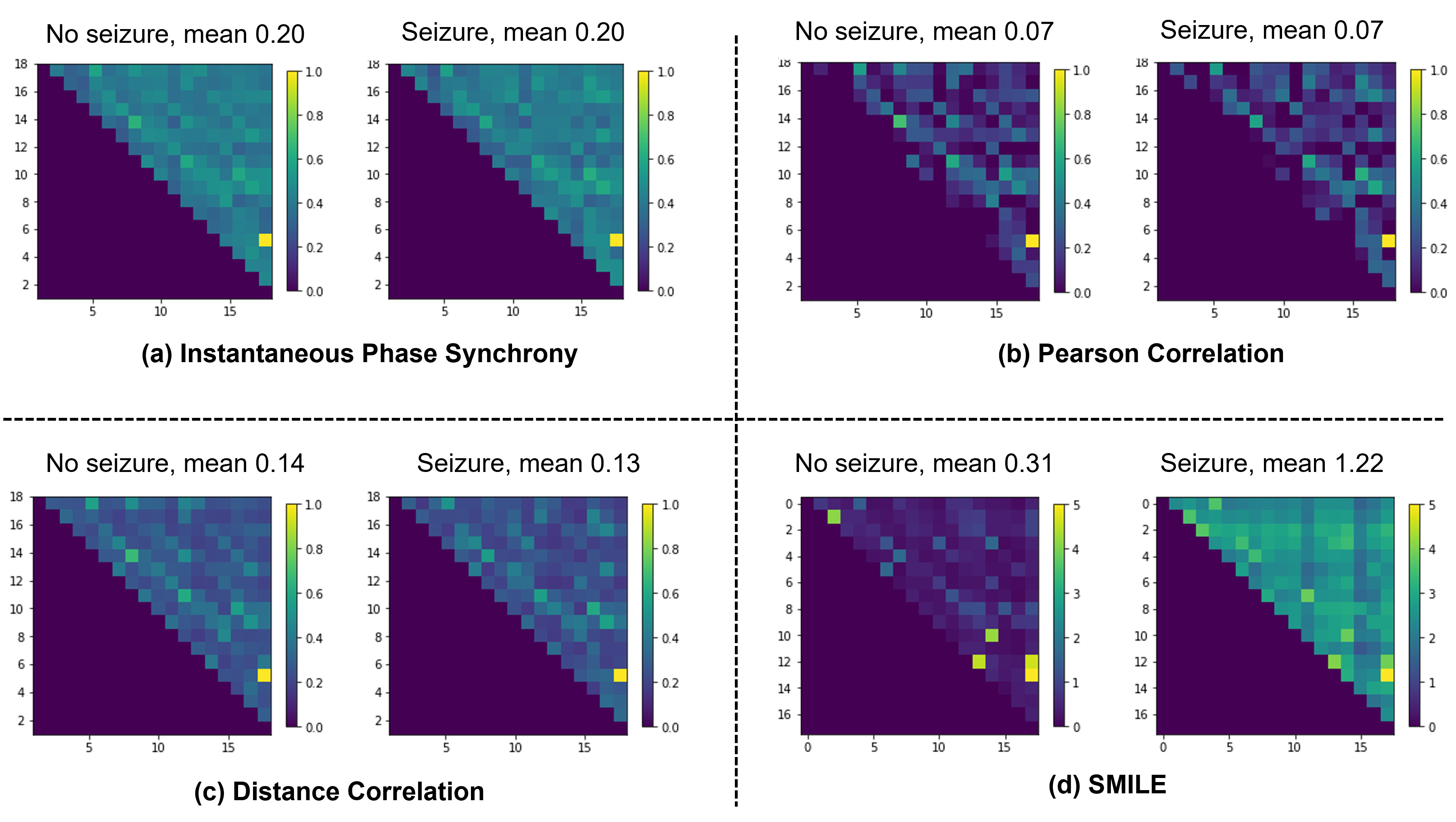}} 
\caption{Four different correlation measures for seizure and non-seizure. (a) instantaneous phase synchrony, (b) Pearson correlation, (c) distance correlation, and (d) SMILE.   
}
\label{fig:mi result}
\end{figure*}

In order to ensure \ac{mi} is a good measure to compute dependency between \ac{eeg} channels, we compare SMILE with three other popular correlation measures:

\begin{itemize}
\item The instantaneous phase synchrony measures the phase similarities between signals at each time-point~\cite{zarghami2020deep}. 

\item The Pearson correlation measures the strength of the linear relationship between two random variables~\cite{benesty2009pearson}.

\item The distance correlation is a measure of association strength between non-linear random variables~\cite{edelmann2021relationships}.
\end{itemize}
Fig.~\ref{fig:mi result} compares these three methods with SMILE. Specifically, we evaluate the degree of dependence between all pairs of channels for time blocks of 4 seconds during seizure as well as during no-seizure regions. We then evaluate average across all blocks for the seizure and no-seizure zones. Based on these mean values, SMILE is the most powerful indicator of highly inter-channel correlation during the seizure compared to the other three correlation measures, which was in agreement with manual inspection of the \ac{eeg} signals.

Based on the above observation, we use SMILE in \ac{mical}  to estimate $I(\boldsymbol{x}_{t}^{(i)};\boldsymbol{x}_{t}^{(j)})$ at each block $t$ for each channel pair $i,j$. Since \ac{mi} is symmetric, i.e., $I(\boldsymbol{x}_{t}^{(i)};\boldsymbol{x}_{t}^{(j)}) = I(\boldsymbol{x}_{t}^{(j)};\boldsymbol{x}_{t}^{(i)})$, we only estimate the \ac{mi} for $j > i$. We set the parametric $T_{\boldsymbol{\theta}}$ to be a fully-connected network with two hidden layers with ReLU activations, and we set $\tau = 0.9$ in the objective~\eqref{eqn:smile}. 
% This is illustrated in Fig.~\ref{fig:mi result}, where  it is observed that the trained estimator outputs higher \ac{mi} values during seizure compared to non-seizure blocks.

% \begin{equation}E=mc^2.\label{eq}\end{equation}

\subsection{1D \ac{cnn}}
\label{sec:1DCNN}

\begin{figure} 
  \centering
  \centerline{\includegraphics[width=\columnwidth]{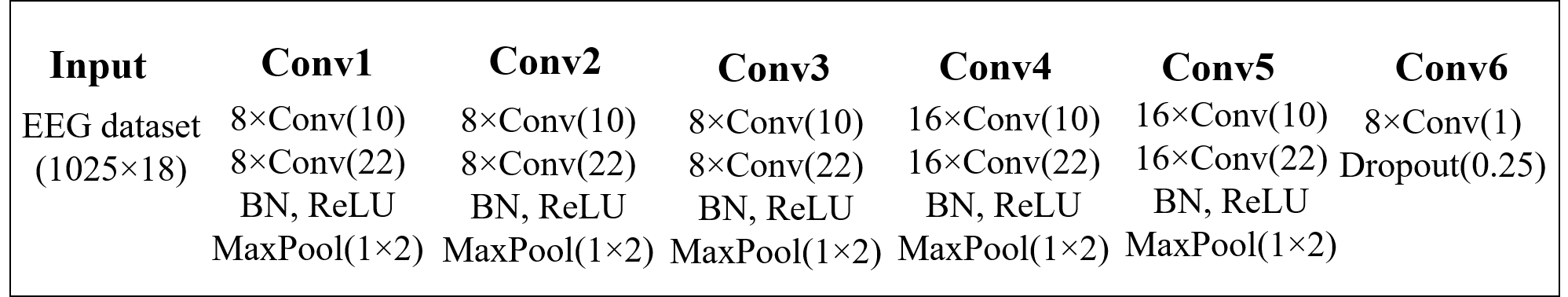}} 
\caption{Proposed 1D CNN architecture. 
%\textcolor{red}{any chance we can make this figure smaller? maybe include only one patioent side by side?}
}
\label{fig:1dcnn}
\end{figure}

In parallel to \ac{mi} estimation, we design and employ a 1D \ac{cnn} in order to produce latent representation of raw \ac{eeg} signals. For having a similar configuration with the baseline model~\cite{boonyakitanont_comparison_2019}, the same number of layers, including convolutions, pooling, dropout, and fully connected layers, are chosen. As shown in Fig.~\ref{fig:1dcnn}, we design the kernel size to obtain a high receptive field compared to prior works. Our proposed 1D \ac{cnn} is able to cover almost 1 second of data, while previous studies had a receptive field of only 30 milliseconds. Having a high receptive field leads to capturing long-term correlation as well as capturing low-frequency components of \ac{eeg} signals. Moreover, compared to 2D \ac{cnn}, 1D \ac{cnn} can operate on all EEG channels at a given time instance. Details of the proposed \ac{cnn} are described in Fig.~\ref{fig:1dcnn}. 

The final set of features that are used for estimating the probability of seizure over a given block is obtained by combining the result of the 1D \ac{cnn} extractor and the estimated \ac{mi}. Specifically, let $\hat{\mathbf{I}}_{t_k}$ be the estimated \ac{mi} values between channel pairs at time $t_k$, and $\mathbf{z}_{t_k}$ be the features extracted by the 1D \ac{cnn}. The the final set of features used for seizure detection is given by $\mathbf{y}_{t_k} = [\hat{\mathbf{I}}_{t_k}, \mathbf{z}_{t_k}]$. This feature is then used as an input to a logistic regression layer for a soft estimate of the seizure event.

%\textcolor{red}{Please add here a paragraph describing how the MI features ad the CNN extracted features are combined into a soft estimate for each EEG block which accounts only for the set of EEG measurements during the considered block. This is where we should introduce the notation of $y$ as the extracted features, not in the following section.}

% \vspace{-0.8cm}

\subsection{Factor Graph Inference}
\label{sec:FG}
Predicting seizure solely based on combined features from \ac{mi} estimator and 1D \ac{cnn} does not take into account past and future \ac{eeg} blocks. Therefore, we utilize the block-wise soft decision not as a direct estimate of the corresponding seizure state, but as learned function nodes in a factor graph incorporating the presence of temporal correlation. Our proposed approach follows the methodology of learned factor graphs utilized for sleep stage detection in~\cite{shlezinger2020learned} and for symbol detection in~\cite{shlezinger2020data}. To describe this operation, we first briefly recall factor graph inference, after which we explain how it is incorporated by \ac{mical} to account for temporal correlation.

% \begin{figure} 
%   \centering
%   \centerline{\includegraphics[width=0.9\columnwidth]{Figures/FGExample_Forney.PNG}} 
% %
% \caption{\textcolor{red}{Remove this image as it does not add anything to the paper}Forney-style factor graph}.

% \label{fig:fgxmp}
% %
% \end{figure}

Factor graphs are a representation of the factorization of local functions of several variables, typically of joint distribution measures, forming a graphical structure~\cite{loeliger2004introduction}.
In the Forney-style factor graph (adopted here) of a joint distribution, the random variables correspond to edges, and their statistical dependence is captured as a node in the graph, referred to as the function node. 
%A factor graph consists of nodes and variables where for every factor and variable, there is unique node and unique edge, respectively. 
As a result, a factor node is connected to a variable edge if and only if the factor is a function of the variable. %Implicitly, no variable appears in more than two factors. 
%As indicated in Fig.\ref{fig:method}, we employ Forney style representation where variable nodes are replaced by edges. 
The key advantage of this representation is that it facilitates extracting quantities which are typically complex to compute, such as marginal probabilities, with a complexity that only grows linearly with the number of variables via, e.g., the sum-product method \cite{kschischang2001factor}. 
To implement factor graph inference, the first step is to create the structure of the graph, i.e., the interconnection between nodes. For this purpose, we use the underlying property of seizure mechanism where the generation of seizure is closely associated with abnormal synchronization of neurons~\cite{mormann2003epileptic}. To incorporate this feature in our model and following consideration \ref{itm:Temporal}, we approximate the temporal relationship as obeying first-order Markovian model. 
%\ac{hmm} is widely used in modeling time-series data, and it is mostly based on the probabilistic method. \ac{hmm} can be represented as a chain model where future and current states are determined by previous state~\cite{lee_classification_2018}. 
%The Markovian architecture leads to exploiting temporal correlation in seizure detection procedures at reduced complexity and makes the model well-suited in real-time applications. 

To formulate this mathematically, the Markovian model implies that the joint distribution of the extracted features $\boldsymbol{y}$ and the latent seizure states $\boldsymbol{s}$ over all $N$ blocks can be factorized as 
%we assume $\boldsymbol{y}=\{y_{1},y_{2},\cdots,y_{N} \}$ as extracted features and latent seizure states $\boldsymbol{s}=\{s_{1},s_{2},\cdots,s_{N} \}$ over $N$ consecutive blocks that can be related by an \ac{hmm}. The latent state supposed to satisfy the Markovian property carries binary values of $s_{i} \in \{0,1\}$, and it shows the presence or absence of seizure in a block. Under \ac{hmm} assumption, the joint probability distribution between extracted features and the states can be formulated as: 
\begin{equation}
    \label{eqn:factor}
	P(\boldsymbol{s},\boldsymbol{y}) =  \prod_{i = 1}^{N} P(s_{i}|s_{i-1}) P(y_{i} | s_{i}).
\end{equation}
Here, $P(s_{i}| s_{i-1})$ represents the seizure state transition probability, which is a control parameter. Based on our numerical experiments, we manually set it to be $89.54\%$ for switching from no-seizure to seizure and $17.9\%$ for opposite situation. The factorization in \eqref{eqn:factor} results in the factor graph representation of the joint distribution $P(\boldsymbol{s},\boldsymbol{y})$ as the sequential graph illustrated in Fig.~\ref{fig:method}

The classification of the sleep states requires to compute marginal distribution $P(s_{i}, \boldsymbol{y})$ from \eqref{eqn:factor}. The sum-product algorithm allows to compute this desirable quantity in a recursive manner via forward and backward message exchanges over the factor graph.
%however, the marginalisation task as a metric to compute the maximum a-posteriori probability detector is computationally expensive since it grows exponentially with the block size. Therefore to overcome this limitation, factor graph inference-based sum-product algorithm is employed to calculate same process linearly. 
In particular, the sum-product method computes the marginal probabilities via
\begin{equation}
\label{eqn:msgs}
    P(s_{k},\boldsymbol{y}) = \mu_{f_{j}\rightarrow s_{k}}(s_{k})\cdot \mu_{f_{j+1}\rightarrow s_{k}}(s_{k}),
\end{equation}
for each $k \in \{1,\ldots,N\}$. 
In \eqref{eqn:msgs}, $\mu_{f_{j}\rightarrow s_{k}}(s_{k})$ is interpreted as forward message 
\begin{equation}
\label{eqn:ForwardPass}
\mu_{f_{j}\rightarrow s_{k}}(s_{k})=\sum_{\{s_{1},\cdots,s_{k-1} \}}\prod_{i = 1}^{n}f_{i}(y_{i},s_{i},s_{i-1}),
\end{equation}
and $\mu_{f_{j+1}\rightarrow s_{k}}(s_{k})$ as the backward message is achieved by 
\begin{equation}
\label{eqn:BackwardPass}
\mu_{f_{j+1}\rightarrow s_{k}}(s_{k})=\sum_{\{s_{k+1},\cdots,s_{N} \}}\prod_{i = n+1}^{N}f_{i}(y_{i},s_{i},s_{i-1}),
\end{equation}
where $f_{i}(y_{i},s_{i},s_{i-1})$ is the function node which is given by 
\begin{equation}
\label{eqn:functionNodes}
 f_{i}(y_{i},s_{i},s_{i-1})= P(s_{i}|s_{i-1}) P(y_{i} | s_{i}). 
 \end{equation}
The resultant marginal distributions \eqref{eqn:msgs} are compared to a predefined threshold of $T=0.7$ for detection. 

According to \eqref{eqn:functionNodes}, implementing sum-product algorithm requires the knowledge of probability distribution $P(y_{i} | s_{i})$. In practice, obtaining this statistical model that relates observations and time series is a highly complex process. Following \cite{shlezinger2020learned}, we use the joint features of \ac{mi} estimator and 1D \ac{cnn} as the soft estimate of probability distributions to learn function nodes in factor graph. Algorithm~\ref{alg:Algo1} summarizes the steps in \ac{mical} seizure detection.

\begin{algorithm}
	\caption{\ac{mical} seizure detection}
	\label{alg:Algo1}
	\textbf{Inputs:}{ SMILE and 1D CNN networks, estimated state transitions,  \ac{eeg} measurements  $\boldsymbol{X}$,  threshold $T$} \\ % COMP threshold $\tau$  }
	\nonl\textbf{Feature extraction:}\\
	\For{$k = 1,\ldots n$}{
	  Apply SMILE to estimate $\hat{I}_{\boldsymbol{\theta}}(\boldsymbol{x}_{t_k}^{(i)};\boldsymbol{x}_{t_k}^{(j)})$, $j > i$\;
	  Apply 1D CNN to obtain combined features $\boldsymbol{y}_{t_k}$ \;
	  Apply dense layer to obtain soft decision $\hat{P}(s_{t_k}|\boldsymbol{y}_{t_k})$\;
	 } 
	\nonl\textbf{Factor graph inference:}\\
	  Compute $\{f_i\}$ from soft decisions via \eqref{eqn:functionNodes}\;
	  \For{$k = 1,\ldots n$}{
	  Compute $\mu_{f_{t_k}\rightarrow s_k}(\{0,1\})$ via \eqref{eqn:ForwardPass}\;
	  Compute $\mu_{f_{t_{n-k+2}}\rightarrow s_{n-k+1}}(\{0,1\})$ via \eqref{eqn:BackwardPass}\;
	 }
	Detect seizure at $t_k$ if $\mu_{f_{t_k}\rightarrow s_k}(1)\mu_{f_{t_{k+1}}\rightarrow s_k}(1) > T$.
\end{algorithm}

\section{Results and Discussion}
\label{sec:eval}
In the following subsections, we describe our experimental study of \ac{mical} for seizure detection\footnote{\href{https://github.com/bsalafia/CNN-Aided-Factor-Graphs-with-Estimated-Mutual-Information-Features-for-Seizure-Detection-MICAL.git}{The source code and hyper-parameters can be found on GitHub}.}. We first explain the data used, performance metrics, evaluation methods, and hyperparameter tuning, in Subsections \ref{ssec:Data}-\ref{ssec:hyper}, respectively. We then present the numerical results along with a discussion in Subsection~\ref{sec:num_results}.

\subsection{Data Description}
\label{ssec:Data}
\subsubsection{\ac{eeg} Data}

% \textcolor{red}{I would move this under results section as the first subsection there.}

The dataset used in this work is publicly available CHB-MIT Database collected at the Children’s Hospital Boston. CHB-MIT consists of scalp \ac{eeg} recordings from pediatric subjects with intractable seizures~\cite{goldberger_physiobank_2000}\footnote{This database is available online at PhysioNet (\url{https://physionet.org/physiobank/database/chbmit/})}. Recordings belong to 24 cases with ages from 1.5 to 22. Each patient contains 9 to 42 EDF files from a single subject. All signals were sampled at 256 frequency with 16-bit resolution and seizure start and end times are labeled. 
% Moreover, the International 10-20 system of EEG electrode position was used for these recordings. As indicated in Fig.\ref{fig:electrode}, letters represent the area of the brain under electrodes e.g. F- Frontal lobe and T - Temporal lobe. Also, Even numbers refer to the right side of the head, and odd numbers denote the left side of the head.
% \begin{figure} 
%   \centering
%   \centerline{\includegraphics[width=0.7\columnwidth]{Figures/Electrodes.png}} 
% %
% \caption{Electrode Placement according to the International 10-20 System\cite{prpa2019brain}}.

% \label{fig:electrode}
% %
% \end{figure}

\subsubsection{\ac{eeg} Pre-processing}
\label{sec:preprocess}
The dataset contains 664 EDF files from all patients. Signals were already annotated by "seizure" and "no-seizure" labels where each case has at least two seizures. Note that for the CHB-MIT database, seizure types are not specified. Unlike prior works, we consider simple pre-processing steps for our proposed algorithm that makes it well-suited for real-time applications. Since seizures duration (from 7 seconds to 753 seconds) compared to overall recording (from 959 seconds to 14427 seconds) is very short and to have a balanced dataset; first, EDf files that include at least one seizure are selected. Each recording is then shortened to 10 times the seizure duration before and 10 times the seizure duration after the seizure. Therefore, there are 20 seconds of non-seizure data for every second of seizure data. 
From \ac{eeg} channels, the 18 bipolar montages are chosen: FP1-F7, F7-T7, T7-P7, P7-O1, FP1-F3, F3-T3, T3-P3, P3-O1, FP2-F4, F4-C4, C4-P4, P4-O2, FP2-F8, F8-T8, T8-P8, P8-O2, FZ-CZ, CZ-PZ. Hereafter, a notch filter is applied to remove 60 Hz line noise from each \ac{eeg} signal. To estimate the probability of seizure over the $t$-th second, the past 32 seconds of recording is used to solve the optimization that estimates \ac{mi}. This window size is large enough to incorporate the correlation among measurements during the ictal state and demonstrates the best results over the dataset. In addition, the past 4 seconds of recording is used as input to the 1D \ac{cnn} for estimating meaningful features from \ac{eeg} blocks. The value of 4 seconds is selected to satisfy a good trade-off between the number of samples in a block and the stationarity of the observed signals over a block. 
\subsection{Evaluation Methods and Performance Metrics}
\label{sec:results}
To evaluate the performance of the models, six following metrics are measured:

\begin{itemize}
    \item \textit{AUC-ROC}: is the area under receiver operating characteristics (ROC) curve, which shows the capability of the model to distinguish between seizure and no-seizure samples.
    
    \item \textit{AUC-PR}: is the area under the precision-recall curve that represents success and failure rates meaning that a high area under the curve shows a low false positive rate and low false-negative rate.
    
       \item \textit{Precision}: intuitively shows the ability of the classifier not to label a sample as positive that is negative. 
   
   \item \textit{Recall}: represents the capability of the classifier to find all the positive samples. 
   
    \item \textit{F1 score}: is a harmonic mean of recall and precision.
    % where the former indicates the proportion of real positive cases that are correctly predicted positive and the latter denotes the proportion of predicted positive cases that are correctly real positive~\cite{powers_evaluation_2020}. 

   \item \textit{Accuracy}: implies the number of correct predictions over the total number of predictions. 
\end{itemize}

In our implementation, we consider three different strategies for training the models: 

\begin{itemize}
    \item \textit{6-fold-leave-4-patient-out evaluation}: is a generalized evaluation pipeline as it creates all the possible training and test sets. In this approach for each fold, 20 patients are considered as the train dataset and 4 different cases are kept out for the test. This approach requires building six models. 
    
    \item \textit{All patient evaluation}: chooses one EDF file from each patient as the test and trains on the remaining files of all cases. Since one of the patients has only two seizure files and each time a different random EDF session is considered as the test, this algorithm creates two models.   
    
    \item \textit{Per patient evaluation}:  selects an EDF file with maximum length as the test and trains the model on the remaining samples for each case separately. This approach creates 24 models for every patient in the dataset. 
    
\end{itemize}

% MICAL is generalized well across all patients as we conduct 6-fold-leave-4-patient-out evaluation. Hence, for each fold, four different patients are kept out for the test.

% The seizure probability prediction is done based on input block through 1D~\ac{cnn} as well as combined features from \ac{mi} estimator and \ac{cnn}. We also add two different structures, including GRU cells and factor graph to the 1D \ac{cnn} features to exploit temporal correlation without incorporating the inter-channel correlation.

\subsection{Hyperparameter Tuning}
\label{ssec:hyper}
As mentioned in Subsection~\ref{ssec:Data}, for every file, the ratio between seizure and no-seizure classes is 1:20. This bias in the training dataset can influence the model performance and entirely ignore the minority class (seizure samples). One approach to address the problem of class imbalance is to randomly duplicate examples from the minority class, which is called oversampling. In our proposed method, the batch size for training all networks is 256. In order to incorporate oversampling approach, we select 128 samples from the non-seizure class and randomly copy 128 samples from the seizure class. 
One of the essential hyperparameters in training a neural network is the learning rate. If the learning rate is too low, the weights will be updated slowly. On the other hand, setting the learning rate to significant values will cause undesirable divergent behavior in the loss function. Therefore, we use Learning Rate Schedules that adjust the learning rate during training by reducing the learning rate according to a pre-defined schedule. Here, we start with $0.0001$, and the factor of $0.5$ reduces it until it reaches $10^{-5}$. We also use ADAM optimizer, and the number of epochs for 6-fold-leave-4-patient-out evaluation and all patient training is 100, and 20 is selected as the number of epochs for per patient training.

\begin{table*}
\begin{center}
{\footnotesize
\begin{tabular}{|c|c c c c c c|}\hline &  AUC-ROC &  AUC-PR &  F1 score & Precision &  Recall &  Accuracy \\
\hline
2D~\ac{cnn}~\cite{boonyakitanont_comparison_2019}& $75.86 \pm 0.08$ & $31.96 \pm 0.16$ & $31.25 \pm 0.11$ & $27.09 \pm 0.1$& $44.96 \pm 0.22$ & $89.08 \pm 0.03$\\
\hline
Spectrogram~\cite{jana_1d-cnn-spectrogram_2020}& $71.13 \pm 0.1$ &$30.45 \pm 0.14$ & $28.73 \pm 0.09$ & $28.68 \pm 0.11$& $36.18 \pm 0.19$ &$89.54 \pm 0.05$\\
\hline
1D CNN (Ours) & $74.98 \pm 0.07$ & $39.6 \pm 0.16$ & $35.28 \pm 0.11$ & $32.46 \pm 0.13$& $44.79 \pm 0.15$ & $89.36 \pm 0.0.06$\\
\hline
1D~\ac{cnn}-GRU (Ours) & $76.65 \pm 0.07$  &  $36.99 \pm 0.15$ & $36.6 \pm 0.11$ & $33.33 \pm 0.17$ & $46.91 \pm 0.17$ &$89.98 \pm 0.04$ \\
\hline
1D~\ac{cnn}-FG (Ours) & $77.13 \pm 0.07$ & $42.04 \pm 0.15$ & $37.53 \pm 0.12$  & $35.45 \pm 0.15$ & $46.13 \pm 0.14$ & $89.73 \pm 0.06$\\
\hline
1D~\ac{cnn}-SMILE (Ours) & $84.25 \pm 0.05$  & $41.8 \pm 0.16$ & $37.28 \pm 0.09$ & $31.56 \pm 0.1$ & $52.65 \pm 0.18$ & $89.21 \pm 0.05$ \\
\hline
1D~\ac{cnn}-SMILE-GRU (Ours) & $81.65 \pm 0.04$ & $40.48 \pm 0.15$ & $37.51 \pm 0.07$  & $\boldsymbol{39.21 \pm 0.12}$ & $43.71 \pm 0.18$ & $\boldsymbol{91.43 \pm 0.04}$\\
\hline
\ac{mical} (Ours) & $\boldsymbol{86.01 \pm 0.05}$ &  $\boldsymbol{44.06 \pm 0.16}$ & $\boldsymbol{38.25 \pm 0.1}$ &$31.05 \pm 0.11$ & $\boldsymbol{57.88 \pm 0.19}$ & $88.21 \pm 0.06$\\
\hline
\end{tabular}
}
\end{center}
\vspace{-0.2cm}
\caption{Summary of results for 6-fold leave-4-patients-out validation}
\label{tab:summary6fold}
\vspace{-0.2cm}
\end{table*}

\begin{table*}
\begin{center}
{\footnotesize
\begin{tabular}{|c|c c c c c c|}\hline &  AUC-ROC &  AUC-PR &  F1 score & Precision &  Recall &  Accuracy \\
\hline
2D~\ac{cnn}~\cite{boonyakitanont_comparison_2019}& $86.8 \pm 0.01$ & $48.35 \pm 0.009$ & $43.9 \pm 0.02$ & $34.55 \pm 0.02$ & $60.19 \pm 0.01$ & $91.45 \pm 0.006$ \\
\hline
Spectrogram~\cite{jana_1d-cnn-spectrogram_2020}& $84.65 \pm 0.02$ & $50.8 \pm 0.07$ & $46 \pm 0.06$ & $38.85 \pm 0.06$ & $56.55 \pm 0.06$ & $92.6 \pm 0.01$ \\
\hline
1D CNN (Ours)& $90.85 \pm 0.01$ & $65.66 \pm 0.02$ & $56.2 \pm 0.02$ & $45.55 \pm 0.03$  & $74.35 \pm 0.005$ & $93.55 \pm 0.006$\\
\hline
1D~\ac{cnn}-GRU (Ours)& $88.20 \pm 0.01$ & $59.69 \pm 0.02$ & $59.19 \pm 0.01$ & $53.84 \pm 0.01$ & $66.14 \pm 0.05$ & $94.95 \pm 0.001$ \\
\hline
1D~\ac{cnn}-FG (Ours)& $93.75 \pm 0.002$ & $73.1 \pm 0.01$ & $64.3 \pm 0.02$ & $55.75 \pm 0.04$ & $76.3 \pm 0.01$ & $95.3 \pm 0.007$ \\
\hline
1D~\ac{cnn}-SMILE (Ours)& $93.75 \pm 0.0004$ & $68.7 \pm 0.01$ & $59.54 \pm 0.01$ & $50.15 \pm 0.02$ & $73.55 \pm 0.01$ & $94.44 \pm 0.004$ \\
\hline
1D~\ac{cnn}-SMILE-GRU (Ours)& $91.75 \pm 0.01$ & $63.84 \pm 0.07$ & $61.1 \pm 0.03$ & $55.69 \pm 0.03$ & $67.6 \pm 0.03$ & $95.19 \pm 0.004$ \\
\hline
\ac{mical} (Ours)& $\boldsymbol{95.6 \pm 0.001}$ & $\boldsymbol{74 \pm 0.01}$ & $\boldsymbol{65.4 \pm 0.02}$ & $\boldsymbol{56.09 \pm 0.03}$ & $\boldsymbol{78.75 \pm 0.01}$ & $\boldsymbol{95.39 \pm 0.005}$ \\
\hline
\end{tabular}
}
\end{center}
\vspace{-0.2cm}
\caption{Summary of results for all patient training}
\label{tab:summaryallcase}
\vspace{-0.2cm}
\end{table*}

% \begin{figure} 
%   \centering
%   \centerline{\includegraphics[width=0.9\columnwidth]{Figures/Oversample.png}} 
% %
% \caption{\textcolor{red}{This figure is not needed. Please remove.}Oversampling method}.

% \label{fig:oversample}
% %
% \end{figure}
\subsection{Numerical Results}
\label{sec:num_results}

% Eight different architectures are implemented for comparison including two baseline models based on deep learning design and feature engineering. In \cite{boonyakitanont_comparison_2019}, A 2D \ac{cnn} architecture is applied to 4-second \ac{eeg} blocks and for the second model, the spectrogram features obtaiend in \cite{jana_1d-cnn-spectrogram_2020} are used as the input for our proposed \ac{cnn}. 
In our experiment, we consider eight different configurations for comparison, including two baseline models since they showed the best results compared to previous studies. One model is related to non-feature-based design, and the other is based on a deep learning feature extraction algorithm. The 2D~\ac{cnn} was employed to raw \ac{eeg} signals in~\cite{boonyakitanont_comparison_2019} and Jana et al.~\cite{jana_1d-cnn-spectrogram_2020} proposed an architecture comprised of spectrogram features and a 1D \ac{cnn}; however, we use our proposed \ac{cnn} model to have the same backbone. To explore the effect of each individual element in \ac{mical} on the performance metrics, a complete ablation study is conducted. Hence, it results in three different situations: 1) ignoring inter-channel and temporal correlations among \ac{eeg} recordings (1D \ac{cnn}), 2) Excluding \ac{mi} estimation results (1D \ac{cnn}-FG), 3) Removing temporal correlations (1D \ac{cnn}-SMILE). As another experiment for comparison, we add GRU to 1D \ac{cnn} and 1D \ac{cnn}-SMILE to ensure factor graph is strong enough to exploit temporal correlations while it reduces the complexity compared to \ac{rnn}s.

Table~\ref{tab:summary6fold} to Table~\ref{tab:summarypercase} summarize the average results of all performance metrics for three evaluation pipelines. Since the oversampling approach is not used for evaluation samples, this dataset is imbalanced. Therefore, the accuracy is not a good measure to investigate the performance of the models, and as shown in the tables, the values for this metric are inconsistent. For instance, in Table~\ref{tab:summary6fold} \ac{mical} shows the lowest accuracy score while it achieves the best results for AUC-ROC, AUC-PR, F1 score, and recall. As represented, although \ac{mical} can exploit inter-channel and temporal correlations, the accuracy score for our algorithm is less than 1D \ac{cnn} and 1D \ac{cnn}-SMILE. As such, in our experiments we mainly observe the performance of other five metrics. 

The average results for 6-fold leave-4-patients-out evaluation are listed in Table~\ref{tab:summary6fold}. We find that compared to the baseline models, our proposed \ac{cnn} architecture improves the AUC-PR and recall by 9\% and leads to a 5\% increase in AUC-ROC, F1 score, and precision. As shown in the table, adding or ignoring the features through the ablation study has not considerably changed the accuracy results. Incorporating temporal correlation for evaluation is conducted via 1D \ac{cnn}-GRU and 1D \ac{cnn}-FG. Unlike 1D \ac{cnn}-GRU that decreases the AUC-PR by 3\%, 1D \ac{cnn}-FG causes 2\% improvement for all performance metrics compared to our \ac{cnn} architecture. This, in fact, implies the strength of the proposed factor graph inference for capturing temporal correlations. The reduction in AUC-PR and F1 score for 1D \ac{cnn}-SMILE compared to 1D \ac{cnn}-FG admits that incorporating only raw \ac{eeg} features and \ac{mi} estimations is not sufficient for detecting the seizure. Compared to baseline results, this is also proved by \ac{mical} that it shows overall the best performance results of 15\% increase in AUC-ROC and AUC-PR, up to 10\% for F1 score and 3\% and 20\% improvement for precision and recall, respectively. Note that 1D~\ac{cnn}-SMILE-GRU represents a higher precision value than \ac{mical} as there is a trade-off between precision and recall. Therefore, \ac{mical} shows the best evaluation result for recall while it achieves the lowest precision score among other models in this study. 
 
All patient training results in Table~\ref{tab:summaryallcase} demonstrate that our proposed \ac{cnn} design achieves almost 15\% improvement compared to the baselines. \ac{mical} in comparison to 1D \ac{cnn}-FG and 1D \ac{cnn}-SMILE, where only inter-channel or temporal correlation is captured indicates the best results. \ac{mical} enhances the performance by almost 10\% for AUC-ROC and 20\% for remaining metrics. We also find that adding GRU to the 1D \ac{cnn}-SMILE leads to decreasing the values for some of the performance metrics. Since the models are training on the limited dataset from only 24 cases, including extra layers could result in overfitting. 
The results for per patient training in Table~\ref{tab:summarypercase} also emphasize that considering both temporal and inter-channel correlations leads to the best performance. Since we average the results across evaluation set from all 24 trained models, and there might be some cases as outliers, GRU-based design works slightly better than \ac{mical} in some of the performance measures.

\begin{table*}
\begin{center}
{\footnotesize
\begin{tabular}{|c|c c c c c c|}\hline &  AUC-ROC &  AUC-PR &  F1 score & Precision &  Recall &  Accuracy \\
\hline
2D~\ac{cnn}~\cite{boonyakitanont_comparison_2019}& $72.82 \pm 0.18$ & $34.72 \pm 0.29$ & $31.77 \pm 0.24$ & $27.63 \pm 0.26$ & $53.77 \pm 0.28$ & $79.82 \pm 0.19$ \\
\hline
Spectrogram~\cite{jana_1d-cnn-spectrogram_2020}& $74.03 \pm 0.19$ & $35.30 \pm 0.22$ & $20.81 \pm 0.21$ & $25.96 \pm 0.3$ & $27.07 \pm 0.29$ & $90.93 \pm 0.06$ \\
\hline
1D CNN (Ours)& $82 \pm 0.19$ & $50.59 \pm 0.33$ & $37.53 \pm 0.29$ & $42.24 \pm 0.34$ & $48.77 \pm 0.35$ & $90.16 \pm 0.1$ \\
\hline
1D~\ac{cnn}-GRU (Ours)& $85.99 \pm 0.12$ & $52.23 \pm 0.35$ & $36.92 \pm 0.34$ & $46.15 \pm 0.4$ & $36.55 \pm 0.33$ & $93.69 \pm 0.08$ \\
\hline
1D~\ac{cnn}-FG (Ours)& $86.17 \pm 0.19$ & $58.07 \pm 0.35$ & $46.09 \pm 0.31$ & $40.47 \pm 0.33$ & $76 \pm 0.27$ & $81.1 \pm 0.21$ \\
\hline
1D~\ac{cnn}-SMILE (Ours)& $88.53 \pm 0.11$ & $61.86 \pm 0.3$ & $52.11 \pm 0.29$ & $51.87 \pm 0.31$ & $61.06 \pm 0.32$ & $92.19 \pm 0.1$ \\
\hline
1D~\ac{cnn}-SMILE-GRU (Ours)& $\boldsymbol{90.83 \pm 0.11}$ & $\boldsymbol{68.87 \pm 0.29}$ & $52.04 \pm 0.34$ & $\boldsymbol{67.32 \pm 0.39}$ & $46.79 \pm 0.34$ & $\boldsymbol{95.69 \pm 0.05}$ \\
\hline
\ac{mical} (Ours)& $90.44 \pm 0.1$ & $66.77 \pm 0.3$ & $\boldsymbol{53.46 \pm 0.31}$ & $50.91 \pm 0.32$ & $\boldsymbol{69.59 \pm 0.32}$ & $88.89 \pm 0.15$ \\
\hline
\end{tabular}
}
\end{center}
\vspace{-0.2cm}
\caption{Summary of results for per patient training}
\label{tab:summarypercase}
\vspace{-0.2cm}
\end{table*}

\section{Conclusions}
\label{sec:conclusion}
In this paper, we have developed \ac{mical} which is a hybrid model-based/data-driven seizure detection algorithm. \ac{mical} enables capturing two essential features in \ac{eeg} recordings such as inter-channel dependency during seizures and temporal correlations. The former is extracted through a neural \ac{mi} estimator and the latter is achieved via factor graph inference. To implement \ac{mical}, we first carefully design a 1D \ac{cnn} to extract features from raw \ac{eeg} signals. Then, soft estimates of joint features from \ac{cnn} and \ac{mi} estimator are used as the learned factor graph nodes to capture temporal correlation at reduced complexity. In this study, we also conduct a comprehensive evaluation strategy as well as a perfect ablation study and \ac{mical} achieves the best results in all scenarios.  

% \clearpage

%\section{references}
\bibliographystyle{IEEEtran}
\bibliography{IEEEabrv,Ref}

\end{document}